\documentclass[twocolumn,aps,prb,showpacs,amsmath]{revtex4}

\usepackage{graphicx}

\begin{document}

\title{
Enhanced Stability of Bound Pairs at Nonzero Lattice Momenta
}

\author{Pavel Kornilovitch}
\email{pavel.kornilovich@hp.com}
\affiliation{
2876 N.W. Audene Drive, Corvallis, OR 97330, USA
}

\date{\today}

\begin{abstract}

A two-body problem on the square lattice is analyzed.  The interaction potential consists 
of strong on-site repulsion and nearest-neighbor attraction.  The exact pairing conditions 
are derived for $s$-, $p$-, and $d$-symmetric bound states.  The pairing conditions are strong 
functions of the total pair momentum {\bf K}.  It is found that the stability of pairs 
{\em increases} with {\bf K}.  At weak attraction, the pairs do not form at the $\Gamma$-point 
but stabilize at lattice momenta close to the Brillouin zone boundary.  The phase boundaries in 
the momentum space, which separate stable and unstable pairs are calculated.  It is found that 
the pairs are formed easier along the $(\pi,0)$ direction than along the $(\pi,\pi)$ direction.  
This might lead to the appearance of ``hot pairing spots" on the $K_x$ and $K_y$ axes.   

\end{abstract}


\maketitle

\section{\label{sec:ones}
Introduction
}

The short coherence length observed in high-tem\-pe\-ra\-tu\-re superconductors has stimulated
theoretical research on real-space pairing on the lattice.  Strong electron-phonon and 
electron-electron interactions coupled with week screening and reduced dimensionality lead 
to formation of preformed bound pairs that Bose-condense and form a superfluid state
\cite{MicRanRob1990,AleMott1994}.  The two interacting particles on a lattice constitute a 
class of exactly solvable quantum mechanical problems.  A number of physical characteristics 
can be obtained in closed analytical forms: the pairing threshold, binding energy, wave function, 
effective radius, and others.  It is essential that the reduction of the two-body Schr\"odinger 
equation to a one-body Schr\"odinger equation occurs differently than in a continuous space.  
The underlying lattice introduces a preferred reference frame, which results in non-trivial 
dependencies of the physical quantities on the total lattice momentum of the pair {\bf K}.  

In the present work we show that bound states on a lattice become {\em more} stable at large 
{\bf K}.  This property is opposite to the Cooper pairs in a BCS superconductor.  
The effect is best seen through the {\bf K}-dependence of the threshold value of the attractive 
potential.  In two dimensions, the threshold is always zero if the potential is purely attractive.  
Therefore, it is essential to consider {\em mixed} repulsive-attractive potentials. (Such 
potentials are also more realistic.) A natural choice is the Hubbard on-site repulsion $U$ 
combined with an attraction $V$ acting on one or more layers of nearest neighbors.  The latter 
mimics the overscreened Coulomb interaction resulted from electron-phonon interactions with 
{\em distant} lattice ions \cite{AleKor2002a,AleKor2002b}.     

General properties of the two-body states on a lattice were reviewed by Mattis \cite{Mat1986}.
Lin gave the first solution for the $s$-symmetric singlet ground state in the $U-V$ model 
\cite{Lin1991}.  In fact, that solution was presented as the low-density limit of the $t-J$ 
model.  Petukhov, Gal\'{a}n, and Verg\'{e}s extended the solution to the $p$-symmetric triplet 
and $d$-symmetric singlet pairs, again in the framework of the $t-J$ model. Kornilovitch found 
pairing thresholds for a family of $U-V$ models, in which attraction extended beyond the first 
nearest neighbors \cite{Kor1995}.  Basu, Gooding, and Leung studied pairing in a $U-V$ model
with anisotropic hopping \cite{BasGooLeu2001}.  Two-body problems were also solved for several 
models with multi-band single-particle spectra \cite{AleTraKor1992,AleKor1993,IvaKorBob1994}. 

Previous studies were mostly confined to the $\Gamma$-point of the pair Brillouin zone. 
(Reference [\onlinecite{PetGalVer1992}] does contain some results for the diagonal and the
boundary of the Brillouin zone.)  In this work we focus on the {\bf K}-dependence of the pairs
properties.  We formulate and solve the pair stability problem in its general form.  We define 
the {\em pairing surface} inside the Brillouin zone, which separates the regions of pair stability 
from the regions of pair instability.  We also show that, in general, the shape of the pairing 
surface is different from the shape of the single-particle Fermi surface.  The difference leads 
to the appearance of pairing ``hot spots'', which are regions in the Brillouin zone where 
{\em real space} pairing takes place while being energetically unfavorable elsewhere.

\section{ \label{sec:two}
The two-particle Schr\"odinger equation
}

In this paper we consider the simplest $U-V$ model on the square lattice, the one that includes 
attraction between the first nearest neighbors only.  For more complex attractive potentials, 
see Ref.~[\onlinecite{Kor1995}].  The model Hamiltonian reads
\begin{equation}
H = - t \sum_{{\bf nm} \sigma} c^{\dagger}_{{\bf m}\sigma} c_{{\bf n}\sigma} 
    + U \sum_{\bf n}  n_{{\bf n} \uparrow} n_{{\bf n} \downarrow}  
    - V \sum_{\bf nm} n_{\bf m} n_{\bf n} .
\label{eq:one}
\end{equation}
Here ${\bf m} = {\bf n} + {\bf l}$ denotes a lattice site that is a nearest neighbor to {\bf n}, 
$n_{{\bf n}\sigma} = c^{\dagger}_{{\bf n}\sigma} c_{{\bf n}\sigma}$, and 
$n_{\bf n} = n_{{\bf n}\uparrow} + n_{{\bf n}\downarrow}$.  The rest of notation is standard. 
We allow one spin-up fermion and one spin-down fermion in the system.  It is convenient to {\em not} 
to fix the permutation symmetry of the coordinate wave function from outset.  In this way, the 
singlet and triplet solutions will be obtained simultaneously.  The system is fully described by 
the tight-binding amplitude $\Psi({\bf n}_1, {\bf n}_2)$ or, equivalently, by its Fourier transform 
$\Phi({\bf k}_1, {\bf k}_2)$.  Fourier transformation of the Schr\"odinger equation with Hamiltonian 
(\ref{eq:one}) yields an equation for $\Phi$:
\begin{eqnarray}
[ E \! & - & \! \varepsilon({\bf k}_1) - \varepsilon({\bf k}_2)] \Phi({\bf k}_1, {\bf k}_2) = 
      U \sum_{\bf q} \Phi({\bf q}, {\bf k}_1 + {\bf k}_2 - {\bf q}) 
\nonumber \\
& - & V \sum_{\bf l} e^{-i {\bf k}_1 {\bf l} }
\sum_{\bf q} \Phi({\bf q}, {\bf k}_1 + {\bf k}_2 - {\bf q}) e^{i{\bf ql}} ,
\label{eq:two}
\end{eqnarray}
where ${\bf l} = (\pm 1,0), (0,\pm 1)$ are the four nearest-neighbor vectors of the square lattice, 
$E$ is the combined energy of the particles, and 
$\varepsilon({\bf k}) = -t \sum_{\bf l} \exp(-i{\bf kl}) = -2t (\cos{k_x} + \cos{k_y} )$ is the 
free single-particle spectrum.  The integral equation (\ref{eq:two}) is solved by introducing five 
functions $\Delta({\bf K})$ that depend only on the total lattice momentum 
${\bf K} = {\bf k}_1 + {\bf k}_2$:
\begin{equation}
\Delta_{(0,0)}({\bf K}) \equiv \sum_{\bf q} \Phi({\bf q}, {\bf k}_1 + {\bf k}_2 - {\bf q}) ,
\label{eq:three}
\end{equation}
\vspace{-0.5cm}
\begin{equation}
\Delta_{\bf l}({\bf K}) \equiv  
\sum_{\bf q} \Phi({\bf q}, {\bf k}_1 + {\bf k}_2 - {\bf q}) e^{i{\bf ql}} .
\label{eq:four}
\end{equation}
The wave function is then expressed as follows:
\begin{equation}
\Phi({\bf k}_1, {\bf k}_2) = \frac{ U \Delta_{(0,0)}({\bf K}) 
- V \sum_{\bf l} \Delta_{\bf l}({\bf K}) e^{- i {\bf k}_1 {\bf l}} }
{E - \varepsilon({\bf k}_1) - \varepsilon({\bf k}_2)} .
\label{eq:five}
\end{equation}
Substituting this solution back into the definitions (\ref{eq:three}) and (\ref{eq:four}) one 
obtains a system of linear equations for $\Delta_{(0,0)}$, $\Delta_{(+1,0)}$, $\Delta_{(-1,0)}$, 
$\Delta_{(0,+1)}$, and $\Delta_{(0,-1)}$.  The determinant of that system defines energy $E$ 
as a function of ${\bf K}$:
\begin{equation}
\left\vert 
\begin{array}{ccccc}
L_{0} - \frac{1}{U} & L_{-q_x} & L_{q_x}      & L_{-q_y}     & L_{q_y}      \\
L_{q_x}  & L_{0} + \frac{1}{V} & L_{2q_x}     & L_{q_x-q_y}  & L_{q_x+q_y}  \\
L_{-q_x} & L_{-2q_x}    & L_{0} + \frac{1}{V} & L_{-q_x-q_y} & L_{-q_x+q_y} \\
L_{q_y}  & L_{-q_x+q_y} & L_{q_x+q_y} & L_{0} + \frac{1}{V}  & L_{2q_y}     \\
L_{-q_y} & L_{-q_x-q_y} & L_{q_x-q_y} & L_{-2q_y} & L_{0} + \frac{1}{V}   
\end{array} 
\right\vert = 0 ,
\label{eq:six}
\end{equation}
where
\begin{equation}
L_p = L_p(E,{\bf K}) \equiv  \sum_{\bf q} \frac{e^{ip}}
{E - \varepsilon({\bf q}) - \varepsilon({\bf K}-{\bf q})} .
\label{eq:seven}
\end{equation}
The eigenvector of the above system, upon substitution in Eq.(\ref{eq:five}), defines the pair wave 
function up to a normalization constant.  

The derived formulae illustrate the general structure of two-body lattice solutions.  The 
spectrum equation is a determinant, with the dimension equal to the total number of sites within 
the interaction range.  In the present model this number is five, one for the Hubbard term and 
four for the nearest-neighbor attraction.  An eigenvector of the corresponding matrix determines 
the two-particle wave function.  The matrix elements are momentum integrals of the ratio of two
energy polynomials $P_{n^2-1}(E)/P_{n^2}(E)$, where $n$ is the number of one-particle energy bands.  
On the square lattice $n=1$, hence the simple form of the integrals $L_p$.  The copper-oxygen
plane is a more complex case with $n=3$.  The corresponding two-body lattice problem was solved 
in Ref.~[\onlinecite{AleKor1993}].    

Analytical investigation of the spectrum and wave function may be difficult, especially in cases 
when matrix elements are not calculable in closed forms.  The pairing threshold is easier to find 
because the determinant needs to be analyzed only at one special energy point.  At the same time, 
the threshold is a convenient and sensitive indicator of pair stability.  The binding condition 
of model ({\ref{eq:one}) will be the prime focus of the rest of this paper.

The spectrum equation (\ref{eq:six}) depends on three parameters: the interaction potentials $U$ 
and $V$, and the total lattice momentum {\bf K} of the pair.  A binding condition is defined as 
a functional relation between $U$, $V$, and {\bf K}, when the energy of a bound state becomes equal 
to the lowest energy of two free particles with the {\em same} combined momentum {\bf K}.  At fixed 
{\bf K}, the minimum energy is realized when both particles have equal momenta ${\bf K}/2$.  This 
is because the kinetic energy of relative motion is zero.  This statement can also be derived 
rigorously by minimizing a sum of two single-particle energies.  The minimum energy is given by 
\begin{equation}
E_{0}({\bf K}) =  2 \varepsilon \left( \frac{\bf K}{2} \right) 
               = -4t \cos{\frac{K_x}{2}} - 4t \cos{\frac{K_y}{2}} .
\label{eq:eight}
\end{equation}
Thus real-space bound states are formed by particles moving parallel to each other.  This should be
compared with the Cooper pairs, which are formed by quasiparticles with opposite momenta. 

Imagine a bound state with energy $E({\bf K})$ just below $E_{0}({\bf K})$, and let 
$E({\bf K}) \rightarrow E_{0}({\bf K})$.  The spectrum equation then defines a relationship between 
$U$, $V$, and ${\bf K}$, which is the pairing condition of interest.  Before attempting general 
solution let us consider the $\Gamma$-point.  At ${\bf K} = (0,0)$ the solution acquires additional 
symmetry, which simplifies analysis of the spectrum equation.  It follows from definitions 
(\ref{eq:seven}) that 
$L_{q_x} = L_{-q_x} = L_{q_y} = L_{-q_y}$, $L_{2q_x} = L_{-2q_x} = L_{2q_y} = L_{-2q_y}$, 
and $L_{q_x+q_y} = L_{q_x-q_y} = L_{-q_x+q_y} = L_{-q_x-q_y}$.  Next, introduced a new basis:
\begin{eqnarray}
 \Delta_{0}  & = & \Delta_{(0,0)}                                                , \nonumber \\
 \Delta_{s}  & = & \frac{1}{4} \: [ \Delta_{(1,0)} + \Delta_{(-1,0)} + \Delta_{(0,1)} + \Delta_{(0,-1)} ] 
                                                                                 , \nonumber \\
 \Delta_{p+} & = & \frac{1}{4} \: [ \Delta_{(1,0)} - \Delta_{(-1,0)} + \Delta_{(0,1)} - \Delta_{(0,-1)} ] 
                                                                                 , \nonumber \\
 \Delta_{p-} & = & \frac{1}{4} \: [ \Delta_{(1,0)} - \Delta_{(-1,0)} - \Delta_{(0,1)} + \Delta_{(0,-1)} ] 
                                                                                 , \nonumber \\
 \Delta_{d}  & = & \frac{1}{4} \: [ \Delta_{(1,0)} + \Delta_{(-1,0)} - \Delta_{(0,1)} - \Delta_{(0,-1)} ] .
\label{eq:nine}
\end{eqnarray}
The new basis functions are eigen-functions of the symmetry operators of the square lattice.  
States of different symmetries separate and matrix (\ref{eq:six}) becomes block-diagonal.  
We consider the blocks separately.  

{\em s-symmetry}.  
The $2 \times 2$ block that mixes $\Delta_0$ and $\Delta_s$ has the form
\begin{equation}
\left( \begin{array}{cc}
L'_{00} - \frac{1}{U}  & L'_{0s}                 \\
L'_{s0}                & L'_{ss} + \frac{1}{V}  
\end{array} \right)
\left( \begin{array}{c}
\Delta_0 \\ \Delta_s
\end{array} \right) = 0 ,
\label{eq:ten}
\end{equation}
where 
\begin{eqnarray} 
 L'_{00} & = & \sum_{\bf q} \frac{1}        {E - 2 \varepsilon({\bf q})} , \nonumber \\
 L'_{s0} & = & \sum_{\bf q} \frac{\cos{q_x}}{E - 2 \varepsilon({\bf q})} , \label{eq:eleven} \\
 L'_{ss} & = & \sum_{\bf q} \frac{2\cos{q_x}(\cos{q_x}+\cos{q_y})}
                                            {E - 2 \varepsilon({\bf q})} , \nonumber 
\end{eqnarray}
and $L'_{0s} = 4 L'_{s0}$.  Substitution $E = E_0 = - 8t$ leads to logarithmic divergence of 
all integrals $L'$.  In order to circumvent this difficulty, add and subtract $L'_{00}$, 
$4 L'_{00}$, and $4 L'_{00}$ from $L'_{s0}$, $L'_{0s}$, and $L'_{ss}$, respectively.  
The differences converge in the limit $E \rightarrow -8t$: $L'_{s0} - L'_{00} = \frac{1}{8t}$, 
$L'_{0s} - 4 L'_{00} = \frac{1}{2t}$, and $L'_{ss} - 4 L'_{00} = \frac{1}{2t}$.  Upon expansion 
of the $2 \times 2$ determinant the $(L'_{00})^2$ terms cancel while the rest yields
\begin{equation}
\left( 2tU - 8tV - UV \right) \cdot L'_{00} - \left( V + 2t + \frac{UV}{8t} \right) = 0 .
\label{eq:twelve}
\end{equation}
The equation is satisfied when the coefficient at the divergent $L'_{00}$ is zero.  This results 
in the binding condition for the $s$-symmetric pair \cite{Lin1991,PetGalVer1992}:
\begin{equation}
V > V_s = \frac{2Ut}{U + 8t} .
\label{eq:thirteen}
\end{equation}
In the weak coupling limit $U, V \ll t$, the threshold reduces to $V_s = \frac{1}{4} U$.  The factor 
4 here is the number of nearest neighbors: the on-site repulsion needs to be 4 times stronger to 
balance the attraction acting on 4 sites at once.  In the limit of infinite Hubbard repulsion, the 
attraction threshold $V_s (U \rightarrow \infty) = 2t$ is finite.  Limited influence of an infinite 
repulsive potential is explained by vanishing of the wave function at ${\bf r}_1 = {\bf r}_2$.  
Without the repulsion, the attractive threshold goes to zero, as expected in two dimensions.  
Thus the Hubbard term raises $V_s$ from zero to $2t$. 

{\em p-symmetry}.  
Two $1 \times 1$ blocks describe two-particle states with $p$ orbital symmetry.  Their energy is 
determined by the equation
\begin{equation}
\sum_{\bf q} \frac{2 \sin^2{q_x}}{E - 2 \varepsilon({\bf q})} + \frac{1}{V} = 0 .
\label{eq:fourteen}
\end{equation}
Unlike the $s$-symmetry case the integral converges at $E = -8t$.  Integration results in the 
binding condition for $p$-symmetric pairs \cite{PetGalVer1992}:
\begin{equation}
V > V_p = \frac{2\pi}{\pi-2}\, t = 5.50 \, t .
\label{eq:fifteen}
\end{equation}
Notice that neither the spectrum nor the binding condition depends on the Hubbard potential $U$.

%
\begin{figure}
\includegraphics[width=8.5cm]{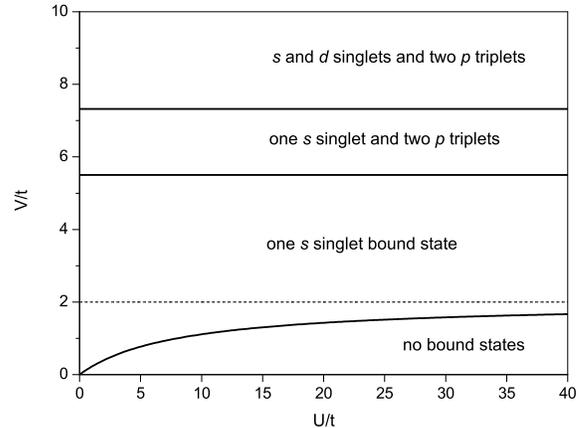}
\vspace{-0.5cm}
\caption{\label{fig:one} 
Phase diagram of the two-particle model (\ref{eq:one}) at ${\bf K} = (0,0)$.  No bound states 
exist below the lowest solid curve.  The phase boundaries are described by Eqs.~(\ref{eq:thirteen}), 
(\ref{eq:fifteen}), and (\ref{eq:seventeen}).  The dashed line indicates the limit value of the 
$s$-state threshold $V_s = 2t$ for an infinitely strong Hubbard potential.  
}
\end{figure}
%

{\em d-symmetry}.
The last $1 \times 1$ block describes two-particle states with $d$ orbital symmetry.  
The spectrum equation is 
\begin{equation}
\sum_{\bf q} \frac{2 \cos{q_x} (\cos{q_x} - \cos{q_y})}
{E - 2 \varepsilon({\bf q})} + \frac{1}{V} = 0 .
\label{eq:sixteen}
\end{equation}
Again, the integral converges in the limit $E = -8t$, which leads to the binding condition 
\cite{PetGalVer1992}:
\begin{equation}
V > V_d = \frac{2\pi}{4-\pi}\, t = 7.32 \, t .
\label{eq:seventeen}
\end{equation}
As with $p$-symmetric pairs, the binding condition does not depend on $U$.  This is because the 
wave functions of the $p$ and $d$ states are identically zero at ${\bf r}_1 = {\bf r}_2$.  The 
contact potential is not felt even at small $V$.

Figure \ref{fig:one} shows the ``phase diagram" of the two-particle system at the $\Gamma$-point.

\section{ \label{sec:three}
The general binding condition
}

Let us return to the full spectrum equation (\ref{eq:six}) and find the binding condition at an 
arbitrary pair momentum ${\bf K}$.  First, using the explicit form of $\varepsilon({\bf q})$ 
and shifting {\bf q} by ${\bf K}/2$ the integrals $L_p$ are transformed as follows:
\begin{eqnarray} 
L_{0}   & = &                     \sum_{\bf q} 
\frac{1}       {E + a \cos{q_x} + b \cos{q_x}}  \equiv                     M_{00}, 
\nonumber \\
L_{q_x} & = & e^{i \frac{K_x}{2}} \sum_{\bf q} 
\frac{\cos q_x}{E + a \cos{q_x} + b \cos{q_x}}  \equiv e^{i \frac{K_x}{2}} M_{10},
\nonumber \\
L_{2q_x} & = & e^{i K_x} \sum_{\bf q} 
\frac{\cos 2q_x}{E + a \cos{q_x} + b \cos{q_x}} \equiv e^{i K_x}           M_{20}, 
\nonumber \\
L_{q_x+q_y} & = & e^{i \frac{K_x+K_y}{2} } \sum_{\bf q} 
\frac{\cos q_x \cos{q_y}}{E + a \cos{q_x} + b \cos{q_x}} 
\nonumber \\
            & \equiv & e^{i \frac{K_x+K_y}{2} } M_{11} ,
\label{eq:eighteen}
\end{eqnarray}
and so on, where 
\begin{eqnarray}
a & \equiv & 4t \cos{(K_x/2)} \geq 0 \nonumber \\
b & \equiv & 4t \cos{(K_y/2)} \geq 0 .
\label{eq:nineteen}
\end{eqnarray}
The subscripts of $M_{nm}$ refer to the frequencies of $\cos{nq_x}$ and $\cos{mq_y}$ in the 
numerator of the integrand.  Upon substitution $E \rightarrow E_{0}({\bf K})$ all $M_{nm}$ 
diverge logarithmically.  Similarly to the $\Gamma$-point case, a divergent $M_{00}$ is 
subtracted and added to each $M_{nm}$.  The differences converge.  A straightforward calculation 
yields:
\begin{eqnarray}
{\bar M}_{10} & \equiv & M_{10} - M_{00} = \frac{2}{\pi a} \arcsin{\sqrt{\frac{a}{a+b}}} ,
\nonumber \\
{\bar M}_{01} & \equiv & M_{01} - M_{00} = \frac{2}{\pi b} \arcsin{\sqrt{\frac{b}{a+b}}} ,
\nonumber \\
{\bar M}_{20} &   =    & \frac{4}{\pi a} \left( 
\frac{a+b}{a} \arcsin{\sqrt{\frac{a}{a+b}}} - \sqrt{\frac{b}{a}} \right) ,
\nonumber \\
{\bar M}_{02} &   =    & \frac{4}{\pi b} \left( 
\frac{a+b}{b} \arcsin{\sqrt{\frac{b}{a+b}}} - \sqrt{\frac{a}{b}} \right) ,
\nonumber \\
{\bar M}_{11} & \equiv & M_{11} - M_{00} = \frac{2}{\pi \sqrt{ab}} .
\label{eq:twenty}
\end{eqnarray}
On the next step, the determinant (\ref{eq:six}) is expanded resulting in a lengthy expression, 
which is in general a fifth-order polynomial in $M_{00}$.  However, the coefficients at the second 
through fifth powers cancel identically.  The spectrum equation reduces to $A \cdot M_{00} + B = 0$, 
where $A$ and $B$ are some complicated functions of ${\bar M}_{nm}$. [$A$ is given below as the 
l.h.s. of Eq.~(\ref{eq:twentyone}).]  Cancellation of the high-order powers of logarithmically 
divergent terms was observed in other two-dimensional models \cite{AleKor1993,IvaKorBob1994,Kor1995}.  
It seems to be a general property of the pairing problem in two dimensions, although no rigorous proof 
is given here.  In the present model, the divergence of $M_{00}$ implies $A = 0$.  After lengthy 
transformations and factorizations the last condition can be written as follows:
\begin{widetext}
\begin{eqnarray}
& \{ & \!\!\!\! 1 - V {\bar M}_{20} \} \cdot \{ 1 - V {\bar M}_{02} \} \cdot
\nonumber \\
& \{ & \!\!\!\! UV^2  \left[ ({\bar M}_{20} - 4 {\bar M}_{10} )({\bar M}_{02} 
 - 4 {\bar M}_{01}) - 4( {\bar M}_{11} - {\bar M}_{10} - {\bar M}_{01} )^2 \right] 
\nonumber \\
& +  & \!\!\! UV \left. \left[ ( {\bar M}_{20} - 4 {\bar M}_{10} ) + 
                    ( {\bar M}_{02} - 4 {\bar M}_{01} ) \right] 
+ V^2 \left[ 8 {\bar M}_{11} - 2 ({\bar M}_{20} + {\bar M}_{02}) \right] 
+ U - 4V \right\} = 0 .
\label{eq:twentyone}
\end{eqnarray}
\end{widetext}
This is the general binding condition sought.  It represents the main analytical result of the 
paper.  The first two factors correspond to the two $p$-symmetric triplet pairs.  This is best 
seen at the diagonal of the Brillouin zone $K_x = K_y$, which implies $a = b$.  This yields 
${\bar M}_{02} = {\bar M}_{20} = (\pi-2)/(2 \pi t \cos{(K_x/2)})$, and the binding conditions 
become
\begin{eqnarray}
V > V_p(K_x = K_y) & = & \frac{2\pi}{\pi-2}\, t \cos{(K_x/2)} \nonumber \\
                   & = & 5.50 \, t \cos{(K_x/2)} .
\label{eq:twentytwo}
\end{eqnarray}
At $K_x = 0$, the last equation reduces to Eq.~(\ref{eq:fifteen}).  Hence, it is identified with 
the $p$-states.   

Equation (\ref{eq:twentytwo}) is a particular example of the general effect: the enhanced stability 
of bound states at large lattice momenta.  Indeed, the pairing threshold {\em decreases} with $K_x$ 
and vanishes completely in the corner of the Brillouin zone.  At any $V < 5.50 \, t$, there exists 
a {\em pairing surface} in the Brillouin zone, which separates stable and unstable bound states 
with $p$ orbital symmetry.  The concrete form of the pairing surface can be found numerically.  

Next, consider the last factor in Eq.~(\ref{eq:twentyone}), which describes the $s$- and 
$d$-symmetric pairs.  The binding condition does not factorize at arbitrary {\bf K}.  However, 
factorization occurs at the diagonal of the Brillouin zone where ${\bar M}_{02} = {\bar M}_{20}$ 
and ${\bar M}_{01} = {\bar M}_{10} = [8 t \cos{(K_x/2)}]^{-1}$
\begin{eqnarray}
s:  & \hspace{0.5cm} & UV ({\bar M}_{20} + 2 {\bar M}_{11} - 8 {\bar M}_{10}) + U - 4V = 0 ,
\nonumber \\
d:  & \hspace{0.5cm} & V ({\bar M}_{20} - 2 {\bar M}_{11}) + 1 = 0 .
\label{eq:twentythree}
\end{eqnarray}
Substitution of the integrals from Eq.~(\ref{eq:twenty}) results in the following binding 
conditions for the $s$ and $d$ states: 
\begin{equation}
V > V_s(K_x = K_y) = \frac{2U \, t \cos{(K_x/2)}}{U + 8t \cos{(K_x/2)}} ,
\label{eq:twentyfour}
\end{equation}
\begin{eqnarray}
V > V_d(K_x = K_y) & = & \frac{2\pi}{4-\pi} \, t \cos{(K_x/2)}  \nonumber \\
                   & = & 7.32 \, t \cos{(K_x/2)} .
\label{eq:twentyfive}
\end{eqnarray}
Comparing the binding conditions on the Brillouin zone diagonal with the $\Gamma$-point, one 
observes that in both cases the hopping integral $t$ is replaced with $t \cos{(K_x/2)}$.  The 
$s$ binding condition approaches $2t \cos{(K_x/2)}$ in the limit of infinite $U$.  The $d$ binding 
condition does not depend on $U$ at all.  Again, the stability of bound states {\em increases} 
with the pair momentum. 

At an arbitrary pair momentum, the pairing surface can be found numerically by solving the 
transcendental equation (\ref{eq:twentyone}).  Figure~\ref{fig:two} shows the pairing surfaces for 
$V = 1.5 \, t$ and $U = 50 \, t$.  If $0 < V < 2Ut/(U+8t)$, there are {\em four} pairing surfaces 
corresponding to the $s$, two $p$, and $d$ pairs.  At $V = 2Ut/(U+8t)$, the $s$ pairing surface 
shrinks to a point indicating that the $s$-symmetrical bound pair is the ground state of the system 
at any momentum {\bf K}.  Similarly, there are two $p$ and one $d$ pairing surfaces at 
$2Ut/(U+8t) < V < 5.50 \, t$, only one $d$ pairing surface at $5.50 \, t < V < 7.32 \, t$, and no 
surfaces at all at $V > 7.32 \, t$.    

\begin{figure}
\includegraphics[width=8.5cm]{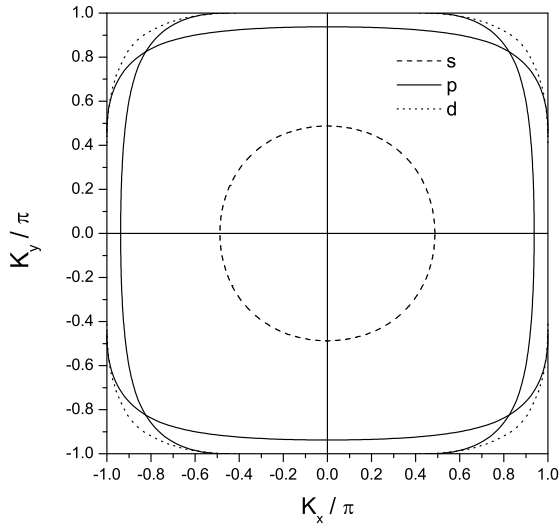}
\vspace{-0.5cm}
\caption{\label{fig:two} 
Pairing surfaces for $U = 50 \,t$ and $V = 1.5 \, t$.  Bound pairs are stable {\em outside} the 
respective surfaces.  At these parameters both $p$ and $d$ surfaces terminate at the Brillouin
zone boundaries: $p$ at $(\pm 0.46 \, \pi , \pm \pi)$ and $(\pm \pi , \pm 0.46 \, \pi)$;
$d$ at $(\pm 0.41 \, \pi , \pm \pi)$ and $(\pm \pi , \pm 0.41 \, \pi)$.  The $p$ and $d$ surfaces 
actually cross.    
}
\end{figure}
\begin{figure}
\includegraphics[width=8.5cm]{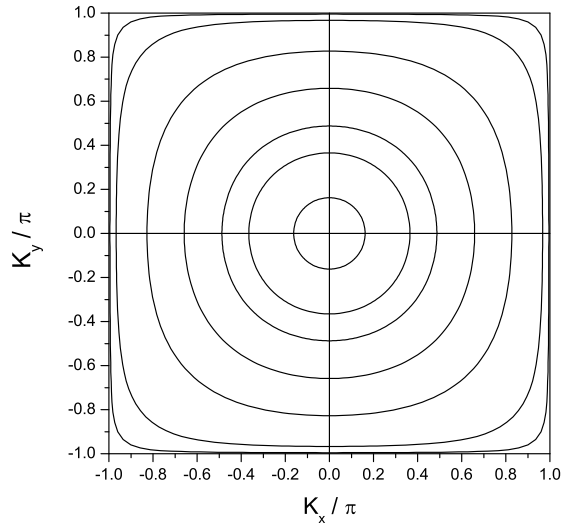}
\vspace{-0.5cm}
\caption{\label{fig:three} 
The pairing surfaces of the $s$-symmetrical bound state for $U = 50 \, t$ and several $V$, obtained
by solving Eq.~(\ref{eq:twentyone}).  Going from outside inward, the plots correspond to 
$V = 0.2\, t$, $0.5 \, t$, $1.0 \, t$, $1.3 \, t$, $1.5 \, t$, $1.6 \, t$, and $1.7 \, t$.  The 
surface shrinks to the origin at $V = 2Ut/(U + 8t) = 1.724 \, t$.  The pairing surface never crosses 
the boundaries of the Brillouin zone.
}
\end{figure}

Let us discuss the evolution of pairing surfaces with $V$ in more detail.  Figure \ref{fig:three} shows 
the evolution of the $s$ surface at $U = 50 \, t$.  At small $V$, the pairing surface approaches the 
boundaries of the Brillouin zone.  For most ${\bf K}$, the strong repulsion overcomes the weak attraction 
and the ground state consists of two unbound particles.  But in the very vicinity of the Brillouin zone 
boundary the scattering phase space shrinks, allowing formation of a bound state with a binding energy 
$\approx V$.  The pair stability region exists for any finite $V$, however small.  The pairing surface 
never crosses the boundary of the Brillouin zone. (The latter fact was noticed in 
Ref.~[\onlinecite{PetGalVer1992}].)  As $V$ increases the stability region expands.  The pairing surface 
shrinks while assuming a circular shape.  Finally, at $V = 2Ut/(U+8t)$ it collapses to the $\Gamma$-point.  

\begin{figure}
\includegraphics[width=8.5cm]{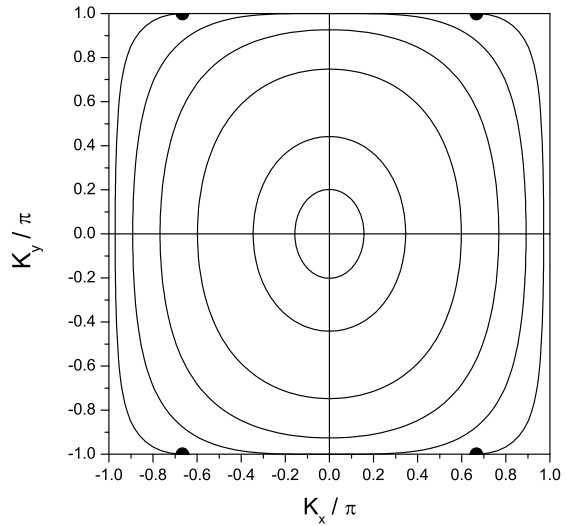}
\vspace{-0.5cm}
\caption{\label{fig:four} 
The pairing surface of one of the two $p$-symmetric bound states for $U = 50 \, t$.  Going from 
outside inward, the lines correspond to $V = 1.0\, t$, $2.0 \, t$, $3.0 \, t$, $4.0 \, t$, $5.0 \, t$, 
and $5.4 \, t$.  The four end points of the $V = 1.0\, t$ plot are indicated by semicircles.  When 
$V \rightarrow 5.50 \, t$, the surface shrinks to the origin but remains elliptic with the 
eccentricity $e = 2\sqrt{(\pi-3)/(3\pi-8)} = 0.63$.  The second family of the $p$ pairing surfaces 
is obtained by rotating the plots by $\pm \pi/2$. 
}
\end{figure}

Figure \ref{fig:four} shows the $V$\!-evolution of one of the $p$-states pairing surfaces.  At small 
$V$ the surface terminates at some points along the $K_y = \pm \pi$ boundaries.  The $K_x$ coordinate 
of the end points is given by the condition $2t \cos{(K_x/2)} = V$.  There are no end points along the 
$K_x = \pm \pi$ boundaries.  At $V = 2t$, the two pairs of end points merge at $(0,\pm \pi)$ resulting 
in an elliptically shaped critical surface.  As $V$ increases further, the critical surface shrinks to 
the origin while retaining the elliptic shape.  The ratio of the two axes of the ellipse is 
$\sqrt{(3\pi-8)/(4-\pi)} = 1.29$.  The pairing surface for the second $p$-symmetric bound state is 
obtained by rotating the plots of Fig.~\ref{fig:four} by $\pm \pi/2$.  

The $d$-state pairing surface is symmetric under the $K_x \leftrightarrow K_y$ exchange, as evident 
from Fig.~\ref{fig:two}.  There are four pairs of end points with the non-trivial momentum coordinate 
determined by the condition $2t(U-2V)\cos{(K_{x,y}/2)} = UV$.  At $V = 2Ut/(U+4t)$, the end points 
merge at $(0,\pm \pi)$ and $(\pm \pi,0)$.  At larger $V$ the pairing surface assumes the circular 
shape similar to the $s$-surface shown in Fig.~\ref{fig:three}.  The $d$-surface shrinks to the origin 
at $V = 7.32 \, t$.

\begin{figure}
\includegraphics[width=8.5cm]{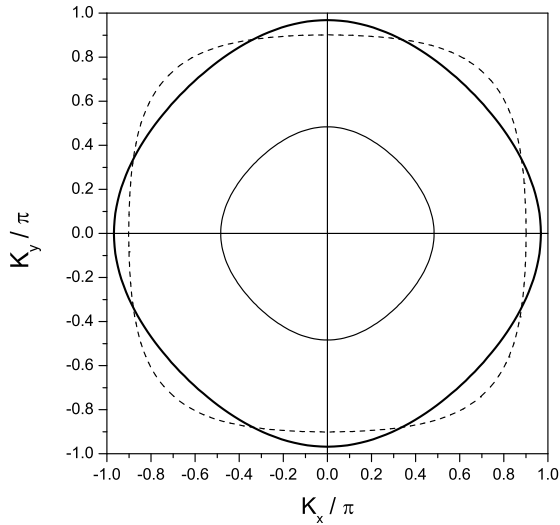}
\vspace{-0.5cm}
\caption{\label{fig:five} 
The pairing surface compared with a single particle Fermi surface.  The thin solid line is the Fermi 
surface defined by the condition $\varepsilon({{\bf k}_F}) = -2.1 \, t$.  The thick solid line is the 
same Fermi surface multiplied by 2.  The dashed line is the $s$-symmetric pairing surface for 
$U = 50 \, t$ and $V = 0.8 \, t$.  Notice the ``binding hot spots'' along the $x$ and $y$ axes where 
the double Fermi surface extends beyond the pairing surface.
}
\end{figure}

\section{ \label{sec:four}
Discussion
}

In continuous space the internal (binding) energy of a composite non-relativistic particle is 
independent of its momentum, and the effective mass is independent of the internal energy.  
In contrast, on a lattice the total mass increases with the internal binding energy \cite{Mat1986}.
In this paper, we established another unusual property of lattice pairs: the stability of a bound
state {\em increases} with its momentum.  Although the total pair energy grows with the momentum, 
the corresponding energy of two free particles grows faster.  As a result, at the Brillouin 
zone boundary the energy of the bound state always lies below the continuum of single-particle 
states.  At the boundary, the periodicity of the pair wave function is commensurate with the 
lattice.  The particles could choose to coherently ``see'' only one component of the potential.  
At the same time, the kinetic energy of the relative motion is suppressed because of shrinking 
of the phase space available for mutual scattering.  Thus as long as the interaction potential 
has at least one attractive region the particles will bind.    

In two dimensions the effect is somewhat hidden by the fact that an arbitrary weak attractive 
potential binds particles even at the $\Gamma$-point.  As {\bf K} increases no qualitative changes 
happen and the pair remains stable.  The situation is different when the potential has strong 
repulsive pieces.  At the $\Gamma$-point the repulsion overcomes the attraction resulting in a 
non-bound ground state.  But at the Brilloiun zone boundary the ground state is always bound.  
By continuity, a pairing surface must exist that separates stable from unstable bound states.  
The shape and size of the pairing surface is a sensitive characteristic of the potential.  
Examples of such surfaces were given in the previous section.

Imagine now a finite density of fermions.  As filling increases, particles with large momenta 
become available.  At some filling, the {\em double} Fermi surface crosses the pairing surface. 
(The single-particle Fermi surface should be multiplied by a factor of 2 for adequate comparison.)  
In general the two surfaces have different shapes.  Therefore crossing begins at some ``hot'' 
segments of the Fermi surface.  The phenomenon is illustrated in Figure \ref{fig:five}.  The 
pairing surface for $U = 50 \, t$ and $V = 0.8 \, t$ is compared with the double Fermi surface 
corresponding to a Fermi energy $-2.1 \, t$ (filling = 0.35 particles per unit cell).  Notice how 
the pairing surface is elongated along the diagonals of the Brillouin zone while the Fermi surface 
is elongated along the $K_x$ and $K_y$ axes.  The overlay of the surfaces creates four regions near 
$(\pm \pi, 0)$, $(0, \pm \pi)$ where pairing correlations are enhanced.  Under such conditions a 
particle with a momentum close to $(\pm \pi/2, 0)$ or $(0, \pm \pi/2)$ can pick up a second particle 
with the {\em same} momentum and form a bound state.  Such a pairing mechanism is in contrast with 
the Cooper pairing.  Obviously, the presence of filled states modifies the low-energy scattering 
dynamics of the pair.  Analysis of the many-body model (\ref{eq:one}) near a paring instability
is a difficult problem, which warrants a separate investigation.  An interesting question is whether
the ``$U$-negative $V$'' model possesses a state where bound pairs with non-zero momenta ``float'' 
atop a fermion sea.  Apart from academic interest, such a state may be of interest for the pseudogap 
phenomenon in high-temperature superconductors. 
(For the latter see, e.g., Ref.~[\onlinecite{DamHusShen2003}].) 
One possible approach to the problem would be the scattering matrix approximation, which has been 
successfully applied to the negative Hubbard model \cite{Mic1995,KyuKleKor1998}.

In conclusion, we have considered a two-particle lattice problem with on-site repulsion and 
nearest-neighbor attraction.  The stability of the bound states {\em increases} with the pair's 
total momentum, which is a consequence of the lattice discreteness.  The effect is conveniently 
described in terms of the pairing surface that separates stable from unstable pairs.  Exact 
pairing surfaces for $s$-, $p$-, and $d$-symmetric bound states have been found and analyzed.  
At finite fillings, the appearance of ``hot pairing spots'' is possible, caused by the difference 
in shapes between the pairing and Fermi surfaces.

\begin{appendix}

\section{\label{app:a}
Bound states in one dimension
}

In this appendix we present, for reference, the two-body solution of model (\ref{eq:one}) is 
{\em one} dimension.  The spectrum equation is a $3 \times 3$ determinant
\begin{equation}
\left\vert 
\begin{array}{ccc}
L_{0} - \frac{1}{U} & L_{-q}              & L_{q}                   \\
L_{q}               & L_{0} + \frac{1}{V} & L_{2q}                  \\
L_{-q}              & L_{-2q}             & L_{0} + \frac{1}{V}  
\end{array} 
\right\vert = 0 ,
\label{eq:aone}
\end{equation}
where
\begin{equation}
L_p = L_p(E,K) \equiv  \sum_{q} \frac{e^{ip}}{E - \varepsilon(q) - \varepsilon(K-q)} ,
\label{eq:atwo}
\end{equation}
and $\varepsilon(k) = -2t\cos{k}$.  The matrix elements are readily calculated.  Expansion of 
the determinant then yields two solutions.

{\em Positive spatial parity (singlet bound state).}  The singlet energy $E_s < 0$ is determined 
from the equation
%
%
\begin{eqnarray}
\left( \sqrt{E^2_s - 16 \, t^2 \cos^2{(K/2)}} + U   \right) \cdot   \makebox[2.0cm]{} \nonumber \\
\cdot \left( \sqrt{E^2_s - 16 \, t^2 \cos^2{(K/2)}} - E_s \right) - \makebox[0.5cm]{} \nonumber \\
- 2 V (U - E_s) = 0  .  
\label{eq:athree}
\end{eqnarray}
%
%
The bound state stabilizes at
\begin{equation}
V > \frac{2U \, t \cos(K/2)}{U + 4t \cos(K/2)} .
\label{eq:afour}
\end{equation}

{\em Negative spatial parity (triplet bound state).}  The triplet energy is given by 
\begin{equation}
E_t = - V - \frac{4 t^2}{V} \cos^2{(K/2)} .
\label{eq:afive}
\end{equation}
The triplet is stable when $V > 2t \cos(K/2)$.  Notice that in the $U \rightarrow \infty$ limit,
$E_s = E_t$.

\end{appendix}


\begin{thebibliography}{99}


\bibitem{MicRanRob1990}
R. Micnas, J. Ranninger, and S. Robaszkiewicz,
Rev. Mod. Phys., {\bf 62}, 113 (1990).

\bibitem{AleMott1994}
A.S. Alexandrov and N.F. Mott,
{\it High Temperature Superconductors and Other Superfluids}
(Taylor \& Francis, 1994).

\bibitem{AleKor2002a}
A.S. Alexandrov and P.E. Kornilovitch, 
J. Phys. Condens. Matter {\bf 14}, 5337 (2002).

\bibitem{AleKor2002b}
A.S. Alexandrov and P.E. Kornilovitch, 
Phys. Lett. A {\bf 299}, 650 (2002).

\bibitem{Mat1986}
D.C. Mattice, 
Rev. Mod. Phys., {\bf 58}, 361 (1986).

\bibitem{Lin1991}
H.Q. Lin,
Phys. Rev. B {\bf 44}, 4674 (1991).

\bibitem{PetGalVer1992}
A.G. Petukhov, J. Gal\'{a}n, and J. Verg\'{e}s,
Phys. Rev. B {\bf 46}, 6212 (1992).

\bibitem{Kor1995}
P.E. Kornilovitch,
in {\it Polarons and Bipolarons in High-T$_c$ Superconductors and Related Materials},
edited by E.K.H. Salje, A.S. Alexandrov, and W.Y. Liang, p. 367 
(Cambridge University Press, 1995).

\bibitem{BasGooLeu2001}
S. Basu, R.J. Gooding, and P.W. Leung,
Phys. Rev. B {\bf 63}, 100506 (2001).

\bibitem{AleTraKor1992}
A.S. Alexandrov, S.V. Traven, and P.E. Kornilovitch,
J. Phys. Condens. Matter {\bf 4}, L89 (1992).

\bibitem{AleKor1993}
A.S. Alexandrov and P.E. Kornilovitch, 
Z. Phys. B {\bf 91}, 47 (1993).

\bibitem{IvaKorBob1994}
V.A. Ivanov, P.E. Kornilovitch, and V.V. Bobryshev,
Physica C {\bf 235-240}, 2369 (1994). 

\bibitem{DamHusShen2003}
A. Damascelli, Z. Hussain, Z.-X. Shen,
Rev. Mod. Phys. {\bf 75}, 473 (2003).

\bibitem{Mic1995}
R. Micnas {\em et al}.,
Phys. Rev. B {\bf 52}, 16223 (1995).

\bibitem{KyuKleKor1998}
B. Kyung, E.G. Klepfish, and P.E. Kornilovitch,
Phys. Rev. Lett. {\bf 80}, 3109 (1998).

\end{thebibliography}
\end{document}